%Paper: hep-ph/9302205
%From: <POIS%TAMPHYS.BITNET@ricevm1.rice.edu>
%Date: Tue, 2 Feb 93 17:03 CST

\input harvmac

\def\Titlehh#1#2{\nopagenumbers\abstractfont\hsize=\hstitle\rightline{#1}%
\vskip .2in\centerline{\titlefont #2}\abstractfont\vskip .2in\pageno=0}
\def\CTPa{\it Center for Theoretical Physics, Department of Physics,
      Texas A\&M University}
\def\CTPb{\it College Station, TX 77843-4242, USA}
\def\HARCa{\it Astroparticle Physics Group,
Houston Advanced Research Center (HARC)}
\def\HARCb{\it The Woodlands, TX 77381, USA}
\def\CERN{\it CERN Theory Division, 1211 Geneva 23, Switzerland}
\def\ie{\hbox{\it i.e.}}

\catcode`\@=11 % This allows us to modify PLAIN macros.

\def\lsim{\mathrel{\mathpalette\@versim<}}
\def\gsim{\mathrel{\mathpalette\@versim>}}
\def\@versim#1#2{\vcenter{\offinterlineskip
    \ialign{$\m@th#1\hfil##\hfil$\crcr#2\crcr\sim\crcr } }}
\def\boxit#1{\vbox{\hrule\hbox{\vrule\kern3pt
      \vbox{\kern3pt#1\kern3pt}\kern3pt\vrule}\hrule}}

\def\t1{{\tilde 1}}
\def\ov{\overline}

\def\eV{\,{\rm eV}}

\def\GeV{\,{\rm GeV}}

\def\NPB#1#2#3{Nucl. Phys. B {\bf#1} (19#2) #3}
\def\PLB#1#2#3{Phys. Lett. B {\bf#1} (19#2) #3}

\def\PRD#1#2#3{Phys. Rev. D {\bf#1} (19#2) #3}
\def\PRL#1#2#3{Phys. Rev. Lett. {\bf#1} (19#2) #3}

\def\MODA#1#2#3{Mod. Phys. Lett. A {\bf#1} (19#2) #3}

\nref\KL{D. A. Kirzhnits and A. D. Linde, \PLB{42}{72}{471}.}
\nref\Weinberg{S. Weinberg, \PRD{9}{74}{3357}.}
\nref\antirest{A. Masiero, T. Yanagida and D. V. Nanopoulos, \PLB{138}{84}{91};
T. W. Kephart, T. J. Weiler and T. C. Yuan, \NPB{330}{90}{705}.}
\nref\Klink{F. R. Klinkhamer and N. S. Manton, \PRD{30}{84}{2212}.}
\nref\Manton{N. S. Manton, \PRD{28}{83}{2019}.}
\nref\KRS{V. A. Kuzmin, V. A. Rubakov and M. E. Shaposhnikov,
\PLB{155}{85}{36}.}
\nref\baryo{M. Shaposhnikov, JETP Lett.{\bf 44} (1986) 465;
\NPB{287}{87}{757}; \NPB{299}{88}{797}; L. McLerran, \PRL{62}{89}{1075}.}
\nref\twodoub{L. McLerran, M. Shaposhnikov, N. Turok and M. Voloshin,
\PLB{256}{91}{451}.}
\nref\DINEa{M. Dine, P. Huet and R. Singleton, \NPB{375}{92}{625}.}
\nref\CKNa{A. G. Cohen, D. B. Kaplan and A. E. Nelson, \PLB{263}{91}{86}.}
\nref\turok{N. Turok and J. Zadrozny, \NPB{369}{92}{729}.}
\nref\SUSY{A. G. Cohen and A. E. Nelson, \PLB{297}{92}{111};
M. Dine, P. Huet, R. Singleton and L. Susskind, \PLB{257}{92}{351};
A. Masiero and A. Riotto, \PLB{289}{92}{73}; G. F. Giudice, A. Masiero,
M. Pietroni and A. Riotto, CERN-TH 6656/92, DFPD 92/TH/43, SISSA-127/92/AP.}
\nref\Mohapatra{R. N. Mohapatra and X. Zhang, \PRD{46}{92}{5331}.}
\nref\Yanagida{M. Fukugita and T. Yanagida, \PLB{174}{86}{45}.}
\nref\CKNb{A. G. Cohen, D. B. Kaplan and A. E. Nelson, \NPB{349}{92}{727}.}
\nref\Dolan{L. Dolan and R. Jackiw, \PRD{9}{74}{3320}.}
\nref\RECENT{M. Dine, R. G. Leigh, P. Huet, A. Linde and D. Linde,
\PRD{46}{92}{550}; C. G. Boyd, D. E. Brahm and S. D. H. Hsu,
CALT-68-1795, HUTP-92-A027, EFI-92-22;
D. E. Brahm and S. D. H. Hsu, CALT-68-1705, HUTP-91-A063;
M. E. Carrington, \PRD{45}{92}{2933}; J. R. Espinosa,
M. Quiros and F. Zwirner, \PLB{291}{92}{115}.}
%\nref\Arnold{P. Arnold, UW/PT-92-06,NUHEP-TH-92-06}
\nref\Grivaz{J.-F. Grivaz, Invited talk at the workshop on `Ten Years of
SUSY Confronting Experiment', CERN, Geneva, 7-9 September 1992, LAL 92-59.}
\nref\Anderson{G. Anderson and L. Hall, \PRD{45}{92}{2685}.}
\nref\Hunters{J. Gunion, H. Haber, G. Kane and S. Dawson, {\it The
Higgs Hunter's Guide}, Addison-Wesley, 1990.}
\nref\Myint{S. Myint, \PLB{287}{92}{325}.}
\nref\Giudice{G. F. Giudice, \PRD{45}{92}{3177}.}
\nref\unific{J. Ellis, S. Kelley and D. V. Nanopoulos, \PLB{249}{90}{44}.}
\nref\CDEOa{B. A. Campbell, S. Davidson, J. Ellis and K. A. Olive,
\PLB{297}{92}{118}.}
\nref\ENO{J. Ellis, D. V. Nanopoulos and K. Olive, UMN-TH-1117/92,
CTP-TAMU-75/92, CERN-TH-6572/92.}
\nref\MSW{L. Wolfenstein, \PRD{17}{78}{2369}; \PRD{20}{79}{2634};
S. Mikheyev and A. Smirnov, Yad. Fiz. {\bf 42} (1985) 1441; Nuo. Cim. {\bf 9C}
(1986) 17.}
\nref\Sher{M. Sher, Phys. Rep. {\bf 179} (1989) 273.}
\nref\Bochkarev{A. I. Bochkarev and M. E. Shaposhnikov, \MODA{2}{87}{417}.}
\nref\aspects{S. Kelley, J. L. Lopez, D. V. Nanopoulos,
H. Pois and K. Yuan, CERN-TH.6498/92,\hfil\break CTP-TAMU-16/92,
UAHEP927, to appear in Nucl. Phys. B.}
\nref\BKSb{A. I. Bochkarev, S. V. Kuzmin and M. E. Shaposhnikov,
\PLB{244}{90}{275}.}
\nref\CW{S. Coleman and E. Weinberg, \PRD{7}{73}{1888}.}
\nref\LNPWZH{J. L. Lopez, D. V. Nanopoulos, H. Pois, Xu Wang and A. Zichichi,
CTP-TAMU-05/93.}
\nref\DL{L. Durand and J. L. Lopez, \PLB{217}{89}{463}; \PRD{40}{89}{207}.}
\nref\ELN{J. Ellis, J. L. Lopez and D. V. Nanopoulos, \PLB{292}{92}{189}.}
\nref\CDEOb{B. A. Campbell, S. Davidson, J. Ellis and K. A. Olive,
\PLB{256}{91}{457}.}
\nref\Arnowitt{R. Arnowitt and P. Nath, CTP-TAMU-65/92,NUB-TH-3055/92.}

\nfig\I{Fig. 1 shows the ratio $R_c\equiv v(T_c)/T_c$
as a function of the SM Higgs mass $m_H$ for $m_t=115\GeV$ (solid line),
$m_t=150\GeV$ (dashed line), and the result for the $2 \over 3$ reduction
of the cubic term for $m_t=115\GeV$ (dotted line).
The critical Higgs mass $m_H\lsim 45\GeV$ can be seen
by imposing $R_c\gsim 1.3$ on the solid line.}

\nfig\II{Fig. 2 shows a scatter plot of $R_c\equiv v_c(T_c)/T_c$ versus
the lightest CP-even $h$ Higgs for 100 distinct minimal supergravity
models for $\xi_0=1,\xi_A=0$ and $\tan\beta=1.2$.
The values for $m_{1/2},m_t$ are
allowed to vary continuously; $m_t\lsim 148\GeV$ due to tree-level
perturbative unitarity
constraints. The circles with crosses correspond to the case
where $m_t=95\GeV$. Increasing $\tan\beta,\xi_0$ causes a systematic downward
shift in the scatter plot. For every point shown, the
corresponding value for the lightest Higgs is $m_h\lsim 32\GeV$, and is
therefore experimentally excluded.}

\Titlehh{\vbox{\baselineskip12pt\hbox{CERN-TH.6763/92}\hbox{CTP--TAMU--87/92}
\hbox{ACT--16/92}\hbox{}}}
{\vbox{\centerline{The Electroweak Phase Transition}
        \vskip2pt\centerline{in Minimal Supergravity Models}}}
\centerline{D.~V.~NANOPOULOS$^{(a)(b)(c)}$ and H. POIS$^{(a)(b)}$}
\smallskip
\centerline{$^{(a)}$\CTPa}
\centerline{\CTPb}
\centerline{$^{(b)}$\HARCa}
\centerline{\HARCb}
\centerline{$^{(c)}$\CERN}
\vskip .1in
\centerline{ABSTRACT}
We have explored the electroweak phase transition in
minimal supergravity models by extending previous analysis of the
one-loop Higgs potential to include finite temperature effects.
Minimal supergravity is
characterized by two higgs doublets at the electroweak scale, gauge coupling
unification, and universal soft-SUSY breaking at the unification scale.
We have searched for the allowed parameter space
that avoids washout of baryon number via unsuppressed anomalous Electroweak
sphaleron processes after the phase transition.
This requirement imposes
strong constraints on the  Higgs sector. With respect to
weak scale baryogenesis, we find that
the generic MSSM is {\it not} phenomenologically acceptable,
and show that the additional experimental and consistency
constraints of minimal supergravity restricts
the mass of the lightest CP-even Higgs even further
to $m_h\lsim 32\GeV$ (at one loop), also in conflict with
experiment. Thus, if supergravity is to allow for baryogenesis via
any other mechanism above the weak scale, it {\it must}
also provide for B-L production (or some other `accidentally' conserved
quantity) above the electroweak scale. Finally, we suggest that the
no-scale flipped $SU(5)$ supergravity model can naturally and economically
provide a source of B-L violation and realistically account for
the observed ratio $n_B/n_\gamma\sim 10^{-10}$.
\bigskip
{\vbox{\baselineskip12pt\hbox{CERN-TH.6763/92}\hbox{CTP--TAMU--87/92}
\hbox{ACT--16/92}}}
\Date{December, 1992}

\newsec{Introduction}

It is now well known that finite temperature effects can considerably
alter the vacuum symmetry of a gauge theory \KL, and can lead to restoration or
anti-restoration \refs{\Weinberg,\antirest} of a particular global or
gauge symmetry due
to the interaction of the theory with a plasma.
In addition, the non-trivial structure of the
electroweak (EW) vacuum naturally leads to the possibility of unsuppressed
baryon number violation at the weak scale via finite temperature
non-perturbative sphaleron transitions \refs{\Klink,\Manton,\KRS}.
It was subsequently suggested that these baryon number violating effects could
actually lead to baryogenesis at the weak scale \baryo,
(${\cal O}(10^2\GeV$)), since the
necessary conditions of (i) C and CP-violation,
(ii) B-violation and (iii) thermal
non-equilibrium could in principle be satisfied.
Recently, several new mechanisms have been proposed which can
apparently account for the observed ratio $n_B/n_\gamma\sim 10^{-10}$.
Some of these mechanisms are rather economical, and involve {\it e.g.}
two higgs doublets \refs{\twodoub,\DINEa,\CKNa,\turok},
supersymmetry \SUSY, left-right models
\Mohapatra, new heavy Majorana
neutrino decays \Yanagida, and a $CP$-violating neutrino mass matrix \CKNb.

In the standard scenario for
the electroweak phase transition involving the decay of the false vacuum,
thermal non-equilibrium requires the
transition to be first order. This crucial requirement has led to a recent
reappraisal of the EW phase transition beyond the classic treatment of
Dolan and Jackiw \Dolan\ and Weinberg \Weinberg. Although there has been
some recent controversy regarding the generic
form of the standard model (SM) scalar
potential, a consensus now seems to have been reached \RECENT, and the higher
order effects considered (to order $\lambda^{3\over 2}$) tame the
infrared divergences and
effectively rescale the cubic term in the scalar potential
by a factor of $2\over 3$. In addition, no linear terms are present.

Irrespective of the details of baryogenesis,
by insisting that unsuppressed sphaleron transitions after
the EW phase transition do not wash out the observed
BAU (Baryon Asymmetry of the Universe), an upper bound to the SM
Higgs mass ensues, since the
quartic coupling is bounded from above. This ensures an adequate
finite-temperature vacuum expectation value at the critical temperature,
$v(T_c)$ so
that the sphaleron transition rates are sufficiently small. In other words, the
sphaleron mass must be sufficiently large ($M_{sph}\sim v(T_c)$) so that
the Boltzmann factor $e^{(-M_{sph}/T)}$ sufficiently
suppresses the transition rate. In the SM, this translates into the limit
$m_H\lsim 45(37)\GeV$ \foot{Hereafter, Higgs mass limits in
parenthesis represent
the $\sqrt{2\over 3}$ reduction due to the higher order effects.}
which is in conflict with the LEP result of $m_H\gsim 60\GeV$
\Grivaz.
In extensions of the SM, the Higgs sector is usually enlarged by including
singlets or additional doublets which generically relax the
SM bound quoted above ($m_H\leq 45\GeV$).
For a general two Higgs-doublet scenario, in order to avoid washout
of $B+L$, the upper Higgs limit
can possibly be as large as $120(98) \GeV$ \turok, and by adding a
singlet \Anderson, the limit $m_H\lsim 150(122)\GeV$ results.

In the minimal supersymmetric model (MSSM), the Higgs sector contains two
complex Higgs doublets. After spontaneous EW symmetry breaking, the physical
Higgs are the $h,H$ (CP-even), $A$ (CP-odd) neutral fields, and the
charged $H^\pm$ Higgs (see Ref. \Hunters\ for a complete discussion).
Now that one-loop corrections to the Higgs masses are
regarded as essential in certain regions of parameter space, the Higgs masses
depend explicitly on the $21$ parameters of the MSSM.
Thus, any constraints to $m_h$ (the lightest $CP$-even Higgs)
at the EW phase transition also depends explicitly on these many
parameters. In order to simplify matters, the allowed
parameter space of a SUSY model
which employs a single SUSY breaking parameter has been
explored in Ref. \Myint. A
more general case has been considered in Ref. \Giudice, however the
conclusions drawn in Refs. \refs{\Myint,\Giudice}
are not necessarily in agreement regarding the upper limit to $m_h$.
Nonetheless, an apparent upper limit to $m_h$ does depend strongly on
$\tan\beta$ and $m_t$. Overall, it appears that $m_h\lsim 65(53)\GeV$,
corresponding to $m_t=200\GeV$ \Myint.
Given the experimental {\it model independent}
LEP limit of $m_h>43\GeV$ \Grivaz, it might appear that
this simplified SUSY model is still
barely viable, and would favor a very heavy top quark with a relatively
light squark spectrum. We argue however that the resulting Higgs spectrum
is very SM-like, and leads to the much more restrictive
limit $m_h\gsim 60\GeV$. It is unlikely that the MSSM is
involved in weak scale baryogenesis.

The consideration of two Higgs doublets at the EW scale has further motivation
in the context of unified models. From the perspective of SUSY unification,
the restriction to two Higgs doublets has been made quite explicit \unific.
By including additional doublets in the theory, the gauge couplings
either fail to unify, or results in a unification scale $M_U$ that leads
to unacceptably fast proton decay \unific.
In this letter we show that if one considers the EW phase transition in
completely realistic
supergravity models, the constraints of unification combined with
the experimental and consistency constraints restricts $m_h\lsim 32(26)\GeV$
in order to avoid a washout of $B+L$ after the EW phase transition.
This is even more restrictive than the  MSSM.
Therefore, unless there is an additional source of $B-L$ production
above the weak scale,
(or possibly some other `accidentally' conserved quantity \CDEOa), baryon
number would not survive, and would be completely
washed out in this class of supergravity unified models.

We therefore present a natural solution to this problem
by considering the flipped $SU(5)$
supergravity model \ENO. The model also possesses two light higgs doublets,
so $B+L$ could also in principle (and most likely will) be washed out.
However, in this scenario out of equilibrium heavy Majorana
neutrino decay provides a natural source
of $\Delta L\not=0$. As $B+L$ is washed out, the sphaleron
transitions effectively process
this $L$-number into a net $B$-number (and ultimately a BAU),
since $B-L$ is conserved during the
transitions. The model therefore naturally connects
a massive neutrino sector (relevant to the MSW solution to the solar neutrino
problem \MSW) to the issue of baryogenesis and provides a comprehensive picture
of solar neutrino physics, dark matter, and baryogenesis.

\newsec{The EW Phase Transition and Weak Scale $\Delta B\not=0$}

Issues related to both weak scale $\Delta B\not= 0$ and the EW phase
transition have been discussed recently and extensively in the literature
\refs{\RECENT,\Anderson}. We confine ourselves
here to a brief discussion of the issues relevant to our calculations and
conclusions.
At finite temperature, the SM scalar Higgs potential takes the
following form at one-loop \Anderson:
\medskip

\eqn\I{V_T(\phi,T)=D(T^2-T^2_c)\phi^2-ET\phi^3+{1\over 4}\lambda_T\phi^4}

\medskip

\noindent
where $\phi$ is the real, neutral component of the scalar Higgs
field and $T_c,E,D,\lambda_T$ are
calculable parameters that depend on the matter content (see Refs. \Anderson,
the first Ref. of \RECENT, or Ref. \Giudice\ for the details). Finally,
$T_c$ is the critical temperature; to a good approximation,
it can be obtained from\foot{This condition effectively determines the
{\it spinodal} point, which is different than the condition
$V_T(T_c,0)=V_T(T_c,v_c)$. The difference is expected
to be small for the range of Higgs masses we consider here \Sher.}

\medskip

\eqn\II{{{\partial^2 V_T(\phi,T_c)}\over {\partial^2 \phi}}|_{\phi=0}
\simeq 0}

\medskip

The finite temperature vacuum expectation value $v(T)$
is obtained from the usual ${{\partial V_T}\over {\partial \phi}}=0$ condition,
and at the critical temperature $T_c$,
\bigskip
\eqn\III{v(T_c)={3ET_c\over \lambda_T}.}
\bigskip
\noindent
We thus see the crucial role of the cubic
term in obtaining
a non-zero $v(T_c)$, necessary for the first-order phase transition where the
true/false vacuum is separated by a barrier.

Due to the non-trivial EW vacuum and the chiral nature of the theory,
unsuppressed,
topology changing $\Delta (B+L)\not=0$ transitions become
possible, particularly at high temperatures, where the probability to go
{\it over} the classical barrier is enhanced enormously.
The essential requirement for ensuring that the baryon number is not washed
out at the weak scale is that the sphaleron transition rate $\Gamma_{sph}<H
\sim e^{-40}T$, where $H$ is the Hubble expansion parameter.
Since $\Gamma_{sph}\sim Te^{-M_{sph}/T}$, this naive analysis implies
$M_{sph}/T_c\gsim 40$. A more detailed calculation \Bochkarev\
shows that baryon number is safe from `washout' provided that
\medskip

\eqn\III{{M_{sph}(T_c)\over T_c}\ge 45\rightarrow
{v(T_c)\over T_c} \equiv R_c\gsim 1.3 .}

\medskip
\noindent
The last inequality is obtained from the specific SM sphaleron solution
\refs{\Klink,\Manton}. From Eqn. (2.3) one can easily see that as
$\lambda_T$ grows, $R_c$ is obviously diminished, thus the zero
temperature Higgs mass $m_H$ has a natural upper limit.
Fig.1 shows the ratio $R_c$, for $m_t=115$ (solid line),
and $150\GeV$ (dashed line). One can observe an overall asymptotic decrease
in $R_c$ with increasing $m_H$, and by requiring $R_c\gsim 1.3$,
the SM limit $m_H\lsim 45\GeV$ is
evident. Also shown in Fig. 1 is the effect of higher order terms (dotted
line) which serve to reduce this limit to $m_H\lsim 37\GeV$.
As we have discussed,
the generic constraint (2.4) has been imposed on extensions of the
standard model and constrains the masses in these extended Higgs sectors. We
now consider the EW constraints on the Higgs sector in minimal
supergravity models.

\newsec{Minimal Supergravity}
Minimal supergravity models can be regarded as $SU(3)\times SU(2)\times U(1)$
models with the minimal three generations and two Higgs doublets of matter
representations at the EW scale (along with superpartners),
and are assumed to unify into a larger gauge group
($SU(5),SO(10)\;{\rm or}\;E6$) at a
unification mass of $M_U\approx10^{16}\GeV$. The five dimensional
parameter space of this model
can be described in terms of three universal soft-supersymmetry breaking
parameters at $M_U$: $m_{1/2},m_0,A$; the top-quark mass $m_t$,
and finally the ratio of Higgs
vacuum expectation values $\tan\beta=v_2/v_1$. We re-define two of
the independent
soft-SUSY breaking parameters as $\xi_0=m_0/m_{1/2}$, $\xi_A=A/m_{1/2}$.
The sign of the superpotential
Higgs mixing term $\mu$ is also undetermined.
The low energy physical masses are then determined from a detailed RG-analysis
of the gauge and Yukawa couplings, the scalar masses and the trilinear A-terms
that all evolve separately from $M_U$ down to $M_Z$. Thus, all of the
low-energy parameters are correlated to GUT-scale parameters
(along with $m_t,\tan\beta$), and the 21 parameters
of a generic, global SUSY analysis are dramatically reduced to five.

Several consistency and
phenomenological constraints restrict the range of the model parameters, such
as the requirement of radiative EW symmetry breaking, a potential
bounded from below, $m^2_{\tilde q},m^2_{\tilde l}>0$
($\tilde q (\tilde l)$ correspond to the squark (slepton) fields)
and the CDF, LEP experimental constraints to $m_{\tilde g,\tilde l,\tilde q},
m_{\tilde \chi^+,\tilde \chi^0},m_{h,A}$. For
a thorough discussion of these relevant details, see Ref. \aspects. After all
of
these constraints have been imposed, one is left with a bounded region in
the $m_t,\tan\beta$ parameter space for a given $\xi_0, \xi_A, m_{1/2}$.
If one makes the choice of a specific model with an underlying
gauge group, the Yukawa relations at
$M_U$ will further constrain the allowed points. In order to make our analysis
here as general as possible, we initially
do not specify the unifying gauge group, however
in our conclusions, we address the consequences with respect to
the specific $SU(5)$ supergravity model.

Regarding the breaking of the EW symmetry, spontaneously broken supergravity
models achieve this goal by inducing radiative corrections to the parameters
in the higgs potential. This has the effect of dynamically
generating vacuum
expectation values for the neutral components of the two higgs doublets.
At zero temperature, the tree-level potential involving the neutral Higgs
fields is \aspects:

\medskip
\medskip
\eqn\I{V_0=(\mu^2+m^2_{H_1})h^2_1+(\mu^2+m^2_{H_2})
h^2_2+2B\mu h_1 h_2+
{{(g^2_2+g'^2)}\over 8}(h^2_1-h^2_2)^2}

\medskip
\medskip
\noindent
where $g'=\sqrt{3\over 5}g_1$ and $g_2$ are the $U_Y(1)$ and $SU_L(2)$
gauge couplings; $h_1=\phi_3,h_2=\phi_7$ are the real, neutral components of
the $H_1,H_2$ complex Higgs doublet fields respectively (and contain the
eight real degrees of freedom $\phi_{i=1,...8}$),
$\mu$ is the Higgs mixing term in the superpotential,
$B$ is a soft-SUSY breaking parameter and finally we require $B\mu<0$
\foot{The normalization conditions for $H_1,H_2$ are chosen so
that the minimum of
$V$ is located at $h_1=v_1,h_2=v_2$; this is in contrast to the
potential in \BKSb. We have compensated for a $1\over \sqrt{2}$ discrepancy
accordingly in Eqn. (2.4).}.
\noindent
Furthermore, $M^2_W={g^2_2\over 2}v^2,
M^2_Z={(g^2_2+g'^2)\over 2}v^2$, and $m_t={1\over \sqrt{2}}
\lambda_tv\sin\beta,m_b=
{1\over \sqrt{2}}\lambda_bv\cos\beta$,
where $\lambda_{t,b}$ are the usual top and bottom quark Yukawa couplings, and
$v^2=v^2_1+v^2_2$ (where $v=246\GeV$).

In the one-loop approximation $V=V_1=V_0+\Delta V$, where \refs{\Sher,\CW}
\medskip
\medskip

\eqn\II{\Delta V={1\over64\pi^2}{\rm STr}\,{\cal M}^4
\left(\ln{{\cal M}^2\over Q^2}-{3\over2}\right)}

\medskip
\medskip

\noindent
in the $\ov {MS}$ scheme, and the supertrace is defined as
${\rm STr}\,f({\cal M})=\sum_j (-1)^{2j}(2j+1){\rm Tr}\,f({\cal M}_j)$.
${\cal M}_j$ are the higgs-field dependent spin $j=0,1/2,1$ mass matrices,
and $Q$ is the renormalization scale.
We obtain the one-loop corrected higgs
boson masses numerically from the standard mass matrix:

\medskip
\medskip

\eqn\III{M_{ij}^2={1\over 2} \left({\partial^2 V\over \partial \phi_i
\partial \phi_j}\right)_{\vev{\phi_3}=v_1,
\vev{\phi_7}=v_2}}

\medskip
\medskip

Initially, all of the parameters in $V$ need to be
specified at zero temperature. For a given point in the
five-dimensional parameter space
$(m_t,\tan\beta,\xi_0,\xi_A,m_{1/2})$ we numerically solve for
$\mu$ and $B$ from
the minimization conditions for the scalar potential; for a given $\tan\beta$
and $M_Z$ at zero temperature (\ie, $v_1(T=0)$ and $v_2(T=0)$)
we find the values of $\mu$ and $B$ which
solve the following conditions:

\medskip
\medskip

\eqn\I{\left({\partial V\over \partial h_{1,2}}\right)_{\vev{h_1}=v_1,
\vev{h_2}=v_2}=0}
\medskip
\noindent
where $V$ is to zero-temperature scalar Higgs potential.
At finite temperature, each fermion/boson of species $i$
adds the following standard term to the effective potential:
\medskip
\medskip
\eqn\IV{\Delta V_T=\eta_i({T^4\over 2\pi^2})F_\pm(y_i),\;
y_i=m_i(h_1,h_2)/T,}
\medskip
\medskip
\noindent
and $\eta_i$ is the multiplicity for each boson/fermion.
In our calculation, we have included the following fields in the supertrace
appearing in Eqn. (3.2): the third generation quarks and squarks
$t,b(12),\tilde t_{1,2}(6)$, $\tilde b_{1,2}(6)$, the gauge bosons $W(6),Z(3)$,
and the Higgs bosons $h,H,A,H^\pm(1)$ (the numbers in parenthesis specify the
multiplicity $\eta_i$).
The functions
$F_-({\rm bosons}),F_+({\rm fermions})$ are given by
the following standard high ($y_i<1$) and low $(y_i>1)$ temperature expansions:
\medskip
\medskip
\eqn\V{F^h_-(y_i)\simeq -{1\over 45}\pi^4+{1\over 12}\pi^2y^2_i-
{1\over 6}\pi y^3_i-{1\over 32}y^4_i Log(y^2_i/c_b)},

\eqn\V{F^h_+(y_i)\simeq -{7\over 8}({1\over 45})\pi^4+{1\over 24}\pi^2y^2_i+
{1\over 32}y^4_i Log(y^2_i/c_f)}
\medskip
\medskip
\noindent
and $Log(c_b)={3\over 2}+2Log(4\pi)-2\gamma_E \simeq 5.41$,
$Log(c_f)={3\over 2}+2Log(\pi)-2\gamma_E\simeq 2.64$ ($\gamma_E$ is the
standard
Euler-Mascheroni constant). Notice the $y^3_i$ infrared
divergent bosonic contribution in Eqn. (3.6).
For the low temperature expansion,
\medskip
\medskip
\eqn\VI{F^l_\pm(y_i)\simeq-\sqrt{{1\over 2}\pi}e^{-y_i}(1+{15\over 8y_i}).}
\medskip
\medskip
\noindent
We have used these expansions, as well as the tenth-order
polynomial expansions for
$F_\pm(y_i)$ given in ref \DINEa\ which is valid for $(1<y_i<3)$;
the results agree to within $10\%$.

In order to find $T_c$, we employ $V_T=V_1+\Delta V_T$ in eqn.(3.3) and
require
\medskip
\medskip
\eqn\VII{Det
\left({\partial^2 V_T\over \partial h_i
\partial h_j}\right)_{\vev{h_1}=0,
\vev{h_2}=0}\simeq 0}
\medskip
\medskip
\noindent
for $i,j=1,2$, in analogy to Eqn. (2.2).
Having determined $T_c$, $V_T(T_c)$ is then minimized with respect to $h_1,
h_2$, using the following conditions in order to
determine $v_1(T_c),v_2(T_c)$:
\medskip
\medskip
\eqn\VIII{
\left({\partial V_T\over \partial h_1}\right)_{{\vev{h_1}=v_1(T_c)}\atop
{\vev{h_2}=v_2(T_c)}},
\left({\partial V_T\over \partial h_2}\right)_{{\vev{h_1}=v_1(T_c)}\atop
{\vev{h_2}=v_2(T_c)}}\simeq 0}
\medskip
\medskip
We have scanned over the $\xi_0,\xi_A,
m_{1/2},\tan\beta,m_t$ parameter space in an effort to find
points for which $R_c>1.3$.
Points in the parameter space which violate this condition are regarded as
baryonically `unstable' at the weak scale. There
are several approximations involved in the calculation which introduce
uncertainties in the
whole procedure at the anticipated $10-20\%$ level, such as the
uncertainty in $T_c$, the
use of the SM sphaleron mass \DINEa, and the neglect of two-loop (and higher)
order EW and QCD effects. Nonetheless, in the following section we
quote specific numerical results, which we therefore consider
to be $\gsim 80\%$ accurate.

\newsec{Results}

In order to gain confidence with our procedure as well as to make contact
with earlier results, we repeated the analysis of Ref. \Myint\ in the case of
a simplified MSSM model, where only degenerate
$\tilde t_{L},\tilde t_{R}$ contributions
along with a single soft-SUSY breaking term was included\foot{For the small
$\tan\beta$ case considered ($\tan\beta\lsim 2$), this
approximation is perfectly adequate.}. We find
numerical agreement of $T_c,m^{max}_h$ to within $10\%$
in a point-by-point comparison; the difference is expected to lie in the
different approximations for $\Delta V_T$ that was used, as well as the
numerical methods employed. In all points considered, the
largest (and acceptable)
values of $R_c$ required very large values of $m_3$ ($m^2_3=\mu B$ for the
potential we consider).
For example, for $\tan\beta=1.52,m_t=115$, $m_3\simeq 1000\GeV$.
This corresponds to $m_h\sim 48\GeV,m_A\sim 1475\GeV$, and a
very SM-like Higgs spectrum, since the $hZZ\;(\sim \sin(\alpha-\beta))$,
$hb\bar b\;(\sim -\sin\alpha/\cos\beta)$ couplings
$\rightarrow 1$ as $m_A\rightarrow \infty$. The preference for this
limit was in fact pointed out in Ref. \DINEa\ for the tree-level situation.
Due to these SM-like couplings,
we expect the much more restrictive SM Higgs experimental limit to apply
here. The reason is the following: for the set of allowed points
in the simplified MSSM model considered in \Myint, we find the value of
$\sin(\alpha-\beta)\gsim 0.99$ with the coupling $h b\bar b>1$. In this case,
$h$ production is not suppressed compared to
the SM, and the experimentally preferred 2-jet signal from the overwhelmingly
dominant $h\rightarrow b\bar b$ will be comparable to the SM (see ref \LNPWZH\
for a more thorough discussion of these details).
Therefore, the present SM analysis should be applicable to the $h$ Higgs,
and the experimental limit $m_h\gsim 60\GeV$ results.
It is therefore unlikely that $B+L$ would survive in this MSSM
model. Therefore, the likelihood that the MSSM alone is involved in weak-scale
baryogenesis is rather remote. The question we consider next is
whether or not these conclusions hold for realistic minimal
supergravity models as well.

For the realistic supergravity case, the parameters are obviously
more constrained. In the case of $m_3$, once the initial values for
$\tan\beta$,
$\xi_0,\xi_A,m_{1/2},m_t$ are given, $\mu,B$ (and thus $m_3$) are determined.
As a result, $m_3$ is {\it not} a free parameter.
We find that the allowed region of $\tan\beta$ for which
$v(T_c)/T_c\ge 1.3$ is even {\it more} restricted than the generic MSSM limit
$\tan\beta\lsim 1.7$.
Fig. 2 shows a scatter plot of $R_c$ versus $m_h$ for
100 distinct minimal supergravity models. We have fixed
$\xi_0=1,\xi_A=0$, $\tan\beta=1.2$;
$m_{1/2},m_t$ are allowed to vary over their allowed values.
For $\tan\beta=1.2$, the tree-level perturbative unitarity constraint
$\lambda_t\lsim 5$ at all scales restricts $m_t\lsim 148\GeV$ \DL.
We find that increasing
$\tan\beta,\xi_0$ drives $R_c$ to smaller values quite rapidly.
Also shown in Fig.2 is a set of points
(circles with crosses) which correspond to
$m_t=95\GeV$. One can see from the figure that near $m_h^{max}\simeq 32\GeV$,
increasing $m_t$ lowers the value for $R_c$, however this shift can be
compensated by decreasing $\xi_0$. This behavior qualitatively reproduces the
result obtained in Ref. \Myint, where increasing values of $m_t$ correspond to
a smaller SUSY breaking scale.
For all points considered,
$m_h$ grows with increasing $m_{1/2}$, but never exceeds $m_h\simeq 32\GeV$.

Changing the value of $\xi_A$ has little effect on the result.
For example, for $\xi_{0,A}=1,0,\;m_{1/2}=70 \GeV, m_t=115 \GeV$,
$R_c\simeq 1.7$. When $\xi_A$ is varied from $0\rightarrow 1$, $R_c$ varies
from $1.7\rightarrow 1.8$.
For the $\mu<0$
possibility, we find that there exists a lower bound for $m_h$ which exceeds
$32\GeV$. Therefore, any possibility for $R_c\gsim 1.3$ is immediately
ruled out in the $\mu<0$ case.
Higher order effects are expected to reduce the upper limit to
$m_h\lsim 26 \GeV$.
Given the fact that the recent LEP experiments restrict $m_h>43\GeV$, it
appears
that a washout of $B+L$ is inevitable at the weak scale. Overall, we
find that $R_c>1.3$ is
only possible for $\tan\beta\lsim 1.3$, and for $\xi_0\lsim 5$. However, even
for this region, $m_h$ is too small, and is experimentally excluded. Therefore,
it is expected that $B+L$ is also washed out in realistic minimal
supergravity models.

\newsec{The Flipped $SU(5)$ Supergravity Scenario}

Recently, one of us (D.V.N.) along with J. Ellis and K. Olive proposed a
natural baryogenesis mechanism that exists in the flipped supergravity
model \ENO,
and utilizes the
Fukugita-Yanagida scenario where heavy Majorana neutrino decay
generates a net $L$-number
which gets processed into $B$-number at the weak scale \Yanagida.
The crucial requirement
is the existence of $\nu^c_i$ which naturally appear in the flipped case, but
is ad-hoc in the minimal $SU(5)$ model.

The generation of a BAU in the flipped model is intimately connected to
the see-saw mechanism for generating neutrino masses
(see Ref. \ENO\ for details), and in the flipped model considered here,
the light neutrino masses are found to correspond roughly to the following
hierarchy: $m_{\nu_\tau}\sim 10\eV, m_{\nu_\mu}\sim 10^{-3}\eV,
m_{\nu_e}\sim 10^{-7}\eV$. Coupled with an anticipated $\sin^2 2\theta_{e\mu}$
mixing between $\nu_e,\nu_\mu$ of the order of $10^{-2}$, this implies that the
model can provide for both an excellent hot dark matter candidate
(the $\nu_\tau$, with $\Omega_{\nu_\tau}\simeq 0.3$), and an acceptable
MSW solution
to the solar neutrino problem with $\Delta m^2\sim 10^{-6}$ \ELN.
For our purposes here regarding the BAU in supergravity models, it was
shown that this same neutrino mass matrix can lead to an
acceptable and {\it natural}
value of $n_B/n_\gamma \sim 10^{-10}$ through out of equilibrium decays of the
$\nu^c$. The final result for $n_B/n_\gamma$
can be expressed in terms of the dilution factor $\Delta$,
the unknown CP-violating phase factor $\delta$, the superheavy
$\nu^c_i$ neutrino masses, the primordial microwave background fluctuations
$\delta \rho/\rho\simeq 5\times 10^{-6}$, and the top quark Yukawa
coupling \ENO:

\eqn\I{{n_B\over n_\gamma}\simeq {9\over 80\pi} |\lambda_{233}|^2
({m_{\nu^c_1} \over m_{\nu^c_3}})\sqrt{{\delta \rho \over \rho}}
{\delta \over \Delta};}

\noindent
by making natural choices for parameters in Eqn. (5.1) (see \ENO\ again for the
relevant details), one finds that

\eqn\II{{n_B\over n_\gamma}\simeq 2\times 10^{-6}{\delta \over \Delta}.}

\bigskip

\noindent
Thus, with a very natural choice of $\delta$, the scheme is completely
consistent with the observed value $n_B/n_\gamma\simeq 3\times 10^{-10}$
(where $\Delta\sim 10^{-3}$). Although $B+L$ is most likely washed out
at the weak scale, prior out of equilibrium heavy neutrino decay allows
the BAU to survive below the weak scale.

\newsec{Conclusions}
The recent developments in baryon number violation at the EW scale may
lead the way to a completely new understanding of the value $n_B/n_\gamma\simeq
10^{-10}$ at a regime possibly within reach of future experimental
probes. Given the
fact that $n_B/n_\gamma\not=0$, any mechanism which hopes to explain this
number
with or without utilizing
the non-trivial, baryon-changing vacuum of the EW sector
must confront the `washout problem', where baryon number can be
enormously reduced after the EW phase transition
if the higgs mass(es) are too heavy, and  `initial conditions' at the weak
scale
after the phase transition dictate that $B-L=0$.
The constraints of gauge coupling unification and proton decay imply
that no more than two doublets can be considered in this type of
analysis. We have
demonstrated here that supergravity models with the
minimal two Higgs doublet structure
and SM particle content (along with superpartners)
at the weak scale generically lead to a washout of $B+L$ below the
EW phase transition,
since $m_h\lsim 32\GeV$, in obvious contradiction with experiment.
Our results here are completely numerical, and a general, analytical treatment
of the two doublet scalar Higgs potential at one-loop (and higher)
perhaps deserves further
study. However, until further progress is made to reduce the
uncertainties inherent in the calculation, this may
be premature.

We therefore believe that we have
explicitly confirmed the suspicion that additional out of equilibrium
$\Delta(B-L)\not=0$
mechanisms (or some other `accidentally' conserved quantity)
are necessary in order to generate a BAU in the context of supergravity
unified models above the weak scale. Although the MSSM does contain additional
sources of CP-violation in the gaugino/higgsino sector,
we argue that supersymmetric baryogenesis
at the weak scale is no longer
possible since the scenario is most likely
already experimentally excluded by SM Higgs searches.
We therefore believe that the mechanism for the BAU resides elsewhere.
In the case of the minimal $SU(5)$ supergravity model
where $\Delta(B-L)=0$, higher dimensional, non-renormalizable interactions
which lead $R$-parity breaking may
be required \refs{\CDEOb,\Arnowitt}; this may arise from as yet unknown
`Planck slop' effects. As a predictive and economical alternative,
we discussed the flipped
supergravity scenario where the observed $n_B/n_\gamma\simeq 3\times
10^{-10}$ can
be naturally accounted for. Baryogenesis originates in the
neutrino sector of the model via out of equilibrium
heavy Majorana neutrino decay. In addition,
the model is found to be quite consistent with the MSW solution to the
solar neutrino problem, and satisfactorily addresses several dark matter
issues.

\bigskip
\bigskip
\bigskip
\noindent{\it Acknowledgments}: This work has been supported in part by DOE
grant DE-FG05-91-ER-40633. The work of D.V.N. has been supported in
part by a grant from Conoco Inc. We wish to thank Jorge Lopez for helpful
comments, and we also wish to thank Stanley Myint for sharing
some of his specific results, as well as several very useful discussions.
\listrefs
\listfigs
\bye